\def\year{2017}\relax
\documentclass[letterpaper]{article} 
\usepackage{aaai17}  
\usepackage{times}  
\usepackage{helvet}  
\usepackage{courier}  
\usepackage{url}  
\usepackage{amsmath}
\usepackage{amssymb}
\usepackage[T1]{fontenc}
\usepackage{graphicx}  
\begin{document}
%
\interfootnotelinepenalty=10000

\title{Listening to the World Improves Speech Command Recognition}
\author{Brian McMahan \and Delip Rao\thanks{Corresponding author}\\
\texttt{(brian|delip)@r7.ai}\\
R7 Speech Sciences Inc\\
San Francisco CA USA\\
}

\maketitle
\begin{abstract}
We study transfer learning in convolutional network architectures applied to the task of recognizing audio, such as environmental sound events and speech commands. 
Our key finding is that not only is it possible to transfer representations from an unrelated task like environmental sound classification to a voice-focused task like speech command recognition, but also that doing so improves accuracies significantly. 
We also investigate the effect of increased model capacity for transfer learning audio, by first validating known results from the field of Computer Vision of achieving better accuracies with increasingly deeper networks on two audio datasets: UrbanSound8k and the newly released Google Speech Commands dataset. 
Then we propose a simple multiscale input representation using dilated convolutions and show that it is able to aggregate larger contexts and increase classification performance. Further, the models trained using a combination of transfer learning and multiscale input representations need only 40\% of the training data to achieve similar accuracies as a freshly trained model with 100\% of the training data.  Finally, we demonstrate a positive interaction effect for the multiscale input and transfer learning, making a case for the joint application of the two techniques. 
\end{abstract}

\section{Introduction}
\begin{table*}[!ht]
\centering
\begin{tabular}{ |l|r|r|r| } 
\hline
 & No. Audio Samples & Total Hours & No. Classes  \\ \hline
 UrbanSound8K & 8732 & 8.75  & 10 \\
 Google Speech Commands & 64721 & 17.97 & 30 \\ \hline
\end{tabular}
\caption{Descriptions of the two labeled audio datasets that were used, UrbanSound8K and Google Speech Commands, in terms of the number of audio samples, total hours, and total number of classes.}
\label{table:dataset}
\end{table*}

Detection of everyday sounds, such as sounds originating from machinery, traffic sounds, animal sounds, and music is essential for building autonomous agents responsive to their surroundings.   
This has myriad applications ranging from autonomous vehicles~\cite{chu06} to surveillance~\cite{ntalampiras09} to monitoring noise pollution in cities~\cite{maijala18,salamon14}.  
Similarly, Spoken Term Recognition~\cite{miller07} has broad applications from conversational agents~\cite{sainath15} to monitoring news~\cite{parlak08}.  
While many approaches have focused individually on the classification of everyday sounds or recognizing speech, there has been little investigation into the relationship between models trained on both tasks.


In this paper, we present a systematic study of several convolutional neural network architectures and their relationship to transfer learning between environmental sounds and speech commands.
Additionally, we introduce a method for increasing the input resolution of the networks using a single layer of dilated convolutions at multiple scales. 
Our experiments are designed to answer questions about model capacity, the effect of multiscale dilated convolutions, and the quality of feature learning on audio spectrograms.

In our first experiment, we study model capacity of several convolutional network architectures by measuring the performance at varying depths, with and without multiscale dilated convolutions as inputs on an environmental sound classification task.

Informed by the first experiment,  we selected a single convolutional network architecture in our second experiment to evaluate the effectiveness of transfer learning from environmental sounds to speech commands. 
Models that were pre-trained on environmental sounds and adapted to speech commands were compared to models trained solely on speech commands.
Additionally, we investigated whether or not multiscale input through dilated convolutions had a significant impact on the transfer learning. 
The results of this experiment strongly suggest that the pre-trained convolutional networks with the multiscale inputs are learning important properties about audio spectrograms.

Finally, in our third experiment, we repeated the second experiment of transfer learning from environmental sounds to speech commands but varied the amount of speech command data used for adapting and training.
We observe that the pre-trained models need far less data to adapt to the new domain and achieved the much higher accuracy than the models trained only on the speech command data. We also report experiments demonstrating that this gain in accuracy and reduction in training data is additive when multi-scale input representations are used with pre-training.

We conclude by discussing the implications and scope of our experiments. 

\section{Related Work}
\label{sec:relatedwork}
Classifying audio signals has a long and diverse history. In particular, the classification of environmental sounds has attracted researchers from speech to signal processing to bioacoustics employing a range of approaches, such as  Support Vector Machines~\cite{temko06}, Random Forests Classifiers~\cite{piczakesc15}, and Multi Layer Perceptrons~\cite{choi16}. 

Recently,~\cite{piczak15} and~\cite{salamon16} show Convolutional Neural Networks outperform the traditional methods. Despite the success, neither of these approaches have investigated extremely deep networks (100+ layers) on audio data, one of the goals of this paper. 
Relatedly, automatic tagging of music has seen several convolutional networks \cite{dieleman2014end,choi2016automatic,lee2017multi}, but the networks have been relatively small compared the ones we investigate in this paper.
In contrast, domains of audio classification have not seen the systematic application of the increasingly deeper convolutional network architectures that have immensely advanced Computer Vision~\cite{deng09,he16,huang16d}.

For audio classification, however, only recently did \citeauthor{hershey17} (\citeyear{hershey17}) apply a 50-layer  Residual Network (also called ResNets)~\cite{he16} and a 48-layer Inception-V3~\cite{szegedy15} network to classify the soundtracks of videos. 
We extend the audio classification task to models larger than a 100 layers. 
Our largest network is a 169 layers deep that we were able to train on a single NVIDIA Titan X GPU in 20 minutes on the UrbanSound8K dataset (~8 hours of training data), without needing any specialized large-scale training infrastructure.

Incorporating information from multiple scales is a challenge to convolutional networks, but recently dilated convolutions have shown efficacy in doing so for image classification tasks~\cite{yu2015multi}. 
Dilations were successfully used by ~\citeauthor{oord16}(\citeyear{oord16}) for a text-to-speech task where the dilated convolution layers are applied hierarchically as a generative model of audio waveforms. 
Previous works on using multiscale spectrogram~\cite{dieleman2014end,choi2016automatic,lee2017multi} do not study the effect of multiscale convolutions on spectrogram features. 
To the best of our knowledge, this is the first work to systematically study the effect of multiple scales of dilated convolutions for audio classification.

A prominent use of convolutional neural networks in Computer Vision is to utilize transfer learning to classify new image categories~\cite{zeiler14}.
We believe this work is the first to investigate transfer learning for deep neural networks with audio inputs and show success on a completely different audio classification task (speech commands vs. environmental sounds). 

\section{Datasets}

In our experiment, we utilize two different datasets. 
The first is a dataset of environmental sounds~\cite{salamon14} called UrbanSound8K.
The second dataset is the recently-released Google Speech Commands~\cite{warden17} dataset. 
Both datasets are collections of audio clips which represent a single class---types of common urban sounds for UrbanSound8K and single word speech utterances for the Speech Commands dataset. 
In our transfer learning experiments, UrbanSound8K serves as the source dataset and the Speech Commands dataset is the target dataset.
\subsection{UrbanSound8K}

The UrbanSound8K dataset, originally derived from the FreeSound\footnote{freesound.org} collection, consists of 8372 audio samples belonging to 10 categories -- \textit{air\_conditioner}, \textit{car\_horn}, \textit{children\_playing}, \textit{dog\_bark}, \textit{drilling}, \textit{engine\_idling}, \textit{gun\_shot}, \textit{jackhammer}, \textit{siren}, and \textit{street\_music}. 
Most audio samples are limited to 4 seconds long. The dataset comes partitioned into 10 folds for cross validation purposes. 
Audio samples are also labeled with their ``salience''---a binary label denoting whether they were recorded in the foreground or background.
While an interesting property, we did not explore how knowledge of salience could be used to improve model performance.
This collection is quite challenging as many of the classes are highly confusable, even to a human ear, like \textit{jackhammer} and \textit{drilling} or \textit{engine\_idling} and \textit{air\_conditioner} due to the high timbre similarity, and the classes \textit{children\_playing} and \textit{street\_music} due to presence of complex harmonic tones. 
The UrbanSound8K dataset was created with a balanced distribution across the classes.

\subsection{Google Speech Commands}

\begin{figure*}[!ht]
\includegraphics[width=\textwidth]{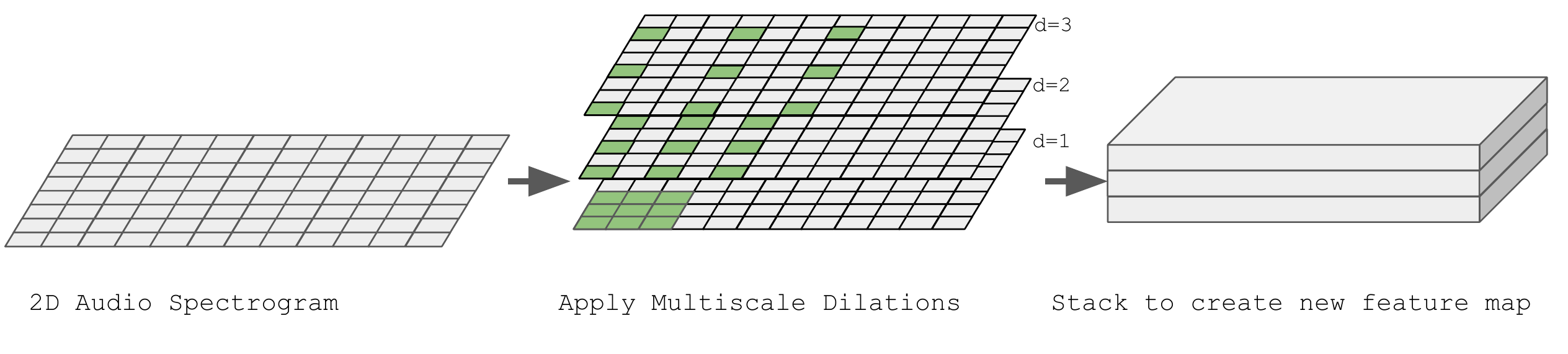}
\caption{Starting with an audio spectrogram, we employ a set of dilations at different scales and with equivalent padding to produce new features maps which can be treated as stacked channels.  Shown here are the first three dilations (d=1, d=2, d=3) with the fourth not shown to save space.}
\label{fig:dilation}
\end{figure*}

The Google Speech Commands dataset is an order of magnitude larger than the UrbanSound8K dataset and completely different in nature (environmental sounds vs. speech). 
The dataset is a crowd-sourced collection of 47,348 utterances of 20 short words — \textit{yes}, \textit{no}, \textit{up}, \textit{down}, \textit{left}, \textit{right}, \textit{on}, \textit{off}, \textit{stop}, and \textit{go}. 
The words also are the class labels to be predicted. 
The utterances are one second long compared to the 4 second samples seen in UrbanSound8K. 
The dataset was gathered by prompting people to speak single-word commands over the course of a five minute session, with most speakers saying each of them five times. 
The dataset also includes 17,373 samples from 10 non-command words. 
These words, like \textit{bed}, \textit{bird}, \textit{cat}, \textit{dog}, \textit{happy}, etc., are unrelated to the core commands and are added to help distinguish unrecognized words. 
Unlike the core command words, the non-command words were said at most once by the speakers. 
Table~\ref{table:dataset} summarizes both datasets.

\section{Feature Extraction}

To prepare the audio data for neural network consumption, each audio file was processed with the following sequence of steps to do minimal feature extraction. 
First, the audio is re-sampled to 22kHz mono and partitioned into overlapping frames.  
The frames are 46 ms long and have 50\% overlap (23 ms).  This is in-line with the feature extraction protocol used by~\citeauthor{salamon16}~(\citeyear{salamon16}). 
Then, the Mel spectrum is extracted using a Fourier transform and a Mel filter bank with 64 filters. 
The preprocessing pipeline is created using the Yaafe audio processing library~\cite{mathieu10}. 
The result of Yaafe preprocessing pipeline is each audio clip transformed into a sequence of frames with each frame being a 64-dimensional feature vector. 
In addition to the Yaafe preprocessing pipeline, we additionally normalize the feature dimensions by subtracting their mean and dividing by their variance (plus a small constant for numeric stability). 
In our experiments, we found this input normalization nonnegotiable. 
Our choice of using the Mel spectral features as opposed to MFCC features commonly used in speech applications is motivated by high noise intolerance of MFCC features.

\section{Models}

In this work, we apply convolutional neural networks to audio spectrograms. 
While seemingly straight forward, there are many methodological considerations that, while explored by the Computer Vision community, have not been as extensively reported for audio spectrograms.
We investigate these considerations by experimentally varying two important properties of modern convolutional networks: network depth using skip connection techniques and multiscale input resolution using dilated convolutions.

\subsubsection{Adapting Convolutional Network Models}
Traditionally, convolutional network architectures have been constructed and validated in Computer Vision settings. 
As a consequence, there are two assumptions that need to changed when adapting image-specific convolutional network models to audio spectrograms.

First is the simpler change of adapting the number of channels in the first-layer to correspond to the single channel mono audio spectrogram.
Second, the networks assume a fixed image size and that images can be cropped or transformed to fit into the fixed size. 
From the fixed size, a combination of striding and pooling is used to reduce the dimensions of the computed tensors until a single vector remains for each data point.
Thus, convolutional networks are adapted to audio spectrograms by replacing the last max pooling layer with one that exactly matches the length and width of the tensor that is output of the network at that point. 

\subsubsection{Baseline Model}

To provide a baseline for the modeling choices presented in this work, we also implement a convolutional network model proposed by~\cite{salamon16}.%
This model, called SB-CNN, is the current state-of-the-art on the UrbanSound8K dataset. 
SB-CNN has three layers of convolutions, interspersed with max-pooling operations. 
It then flattens the third layer's output and applies two fully connected layers.
SB-CNN's design is very similar to traditional feed forward convolutional networks and is a useful comparison for the other models in this study. 

\subsection{Increasingly Deeper Convolutional Networks}
\label{sec:verydeepconvs}

Several techniques have surfaced in recent years which enable dramatically deeper convolutional neural networks.
In this study, we investigate the effectiveness of these techniques on classifying audio spectrograms.
Specifically, we use two architectures, Residual Networks (ResNets) and DenseNets, which employ different techniques to achieve network depth. 

\subsubsection{ResNets} Building on the traditional feed-forward architecture, ResNets~\cite{he16} add a residual connection that allows the output of one layer to skip one or more layers before being summed with the output of another layer.  So, the output of a layer can theoretically depend on the output of all the previous layers and not just the preceding layer.
More formally, let $F_l$ represent the computation of a layer at depth $l$ and $x_{l-1}$ represent the output of computation at layer $l-1$.  
Then, the traditional feed-forward network performs a sequence of operations such that $x_l=F_l(x_{l-1})$.  
With ResNets, a skip connection is added so that the computation of $x_{l-1}$ is summed with the computation of $F_l(x_{l-1})$: 

\begin{equation}
x_l~=~F_l(x_{l-1})~+~x_{l-1} \label{eq:resnet}
\end{equation}

\subsubsection{DenseNets} In addition to ResNets, we utilize a convolutional network architecture named DenseNets due to its state of the art performance and novel use of skip connections. 
More specifically, DenseNets~\cite{huang16d} seized upon a simple observation: convolutional networks greatly benefit from shorter connections between layers closer to the input and layers closer to the output. 
More formally, the computation for $x_l$ is dependent on the computations of all previous layers: 

\begin{equation}
x_l~=~F_l(x_{l-1},~x_{l-2},~\ldots,~x_{0}) \label{eq:densenets}
\end{equation}

Thus, all downstream layers have direct access to the feature maps of all earlier layers.

\subsection{Dilated Kernels for Multiscale Inputs}

Convolutional operations, intuitively, are windowed operations that scan over an input tensor. 
The free parameters of these operations are the size of the window (called the \textit{kernel size}) and the step size of the scan (called the \textit{stride}). 
The kernel size and stride parameterize the receptive field of the convolution: they control how much each convolution operation ``sees'' which allows for designing certain types of information flow. 

More recently, a parameter referred to as \textit{dilation} has been introduced as a way to increase the receptive field without increasing the number of parameters of the convolutional kernel 
\cite{yu2015multi}. 
A dilation is, intuitively, a stride in the kernel---it is a spacing between the scalars in the kernel such that when it is scanned across an input tensor, the kernel subsamples a wider range of input values. 
This is visualized in Figure \ref{fig:dilation}. 
More formally, consider a single position in the output tensor, $Y_{m,n}$.
A convolution operation computes this value by summing over element-wise multiplications, as shown in Equation \ref{eq:basicconv}.
A dilated convolution, however, has a strided kernel such that the positions in the input tensor are spaced further apart.
This is shown in Equation \ref{eq:dilatedconv}.

\begin{align}
Y_{m,n}~&=~\sum^k_{i=0}~\sum^k_{i=0}
  W_{i,j} * X_{m+i,~n+j} \label{eq:basicconv} \\
Y_{m,n}~&=~\sum^k_{i=0} \sum^k_{j=0} 
  W_{i,j} * X_{m+i*d,~n+j*d} \label{eq:dilatedconv}
\end{align}

Using these dilated convolutions, we designed a simple input adapter to existing network architectures.
Specifically, we combine the outputs of four convolutional kernels with dilations of 1, 2, 3, and 4, a kernel size of 3 (both width and height are 3), and a stride of 1. 
The multiscale dilations are used in conjunction with an equivalent padding operation\footnote{If a 3x3 kernel is dilated with $d=2$, then $padding=2$ ensures the output tensor is the same size as the input tensor}, allowing the resulting output tensors to be stacked along the channel dimension.

\section{Experiments and Results}

To evaluate the hypothesized benefit of skip connection convolutional networks and multiscale dilations on audio spectrograms as network inputs, we formulated a series of experiments which varied these two properties. 
More specifically, we first trained the ResNet, DenseNet, and SB-CNN convolutional network architectures on the UrbanSound8K dataset, both with and without multiscale dilations to augment the input audio spectrogram.
These networks were first evaluated directly on the UrbanSound8k dataset by measuring their performance on a held-out test set in a 10-fold cross validation setting.
Then, we selected one of the convolutional network architectures---DenseNet-121---and evaluated whether it was learning information specific only to environmental sounds, or if the learned convolutions would generalize to discriminating voice commands in the Google Speech Commands dataset. 

\subsection{Convolutional Networks on Environmental Sounds}

Our goal is to evaluate whether the choice of convolutional network architecture has an effect on modeling audio spectrograms.
To accomplish this, we utilize the environmental sound classification task of the UrbanSound8K dataset.

\subsubsection{Training Details}

For each of our experimental conditions, we implement the 10-fold cross validation setup provided within the UrbanSound8K dataset~\cite{salamon14}. 
The setup dictates that, for each of the 10 folds, the model is trained on 8 of the folds and at the end of every epoch, validated on the 9th. 
The remaining 10th fold is held out as the final test fold. 
Instead of training for a fixed number of epochs, however, we use an early stopping mechanism that terminates training when performance on the validation fold has not improved for 10 epochs\footnote{We tried different values of this early stopping hyper parameter and found that 10 epoch was a long enough wait that models typically converged.}.  
When the training terminates, the model parameters from the best performing epoch---as measured on the 9th fold---are reloaded and used to evaluate final test set performance. 

\begin{figure*}
\includegraphics[width=\textwidth]{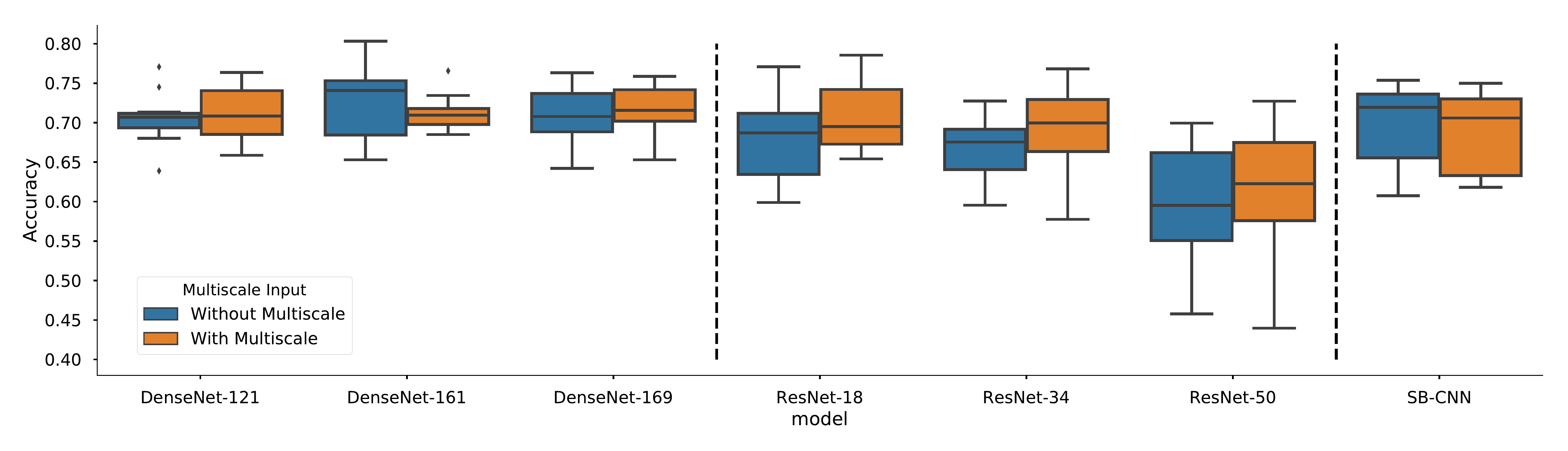}
\caption{The accuracy for the UrbanSound8k dataset---aggregated over 10-fold cross validation---is shown for each of the convolutional network architectures, both with and without multiscale input using dilated convolutions.  The body of each box plot denotes the 25th and 75th percentiles, the line in the body is the median, and the whiskers mark the most extreme observations.}
\label{fig:us8kresults}
\end{figure*}

\subsubsection{Results}

\begin{table}[!tp]
\centering
\begin{tabular}{ |c|c|c| } 
 \hline
  & Without Multiscale & With Multiscale \\
  \hline
  SB-CNN & 69.45 & 68.74 \\
  ResNet18 & 68.12 & 70.92  \\
  ResNet34 & 66.57 & 69.24  \\
  ResNet50 & 59.76 & 61.19  \\
  
  DenseNet121 & 70.61 & 71.04  \\
  DenseNet161 & 72.53 & 71.27  \\
  DenseNet169 & 70.87 & 71.86  \\
   \hline
\end{tabular}
\caption{For each of our independent controls---CNN Architecture and whether to use dilated multiscale input or not---the accuracy on the UrbanSound8K dataset is shown.  The accuracy is the average accuracy from each of the 10-Fold iterations.}
\label{table:us8kresults}
\end{table}

The results are presented in Table \ref{table:us8kresults} and Figure \ref{fig:us8kresults}, but can be summarized with the following observations.
First, the DenseNet architectures---which have been shown to be state-of-the-art for Computer Vision tasks---handily out perform ResNet and the SB-CNN baseline.
Next, ResNet did not perform as well as expected and did not out perform SB-CNN on most comparisons. As a side note, despite our careful effort in closely following the description in~\citeauthor{salamon16} (\citeyear{salamon16}), we were unable to reproduce the SB-CNN accuracy of 73\% as reported by the authors\footnote{Our code, experiment notebooks, library versions, environment details, and hyperparameter setting details are available at: \textsc{URL withheld for blind review.}}.
Finally, there is a general trend of an increase in performance with the use of multiscale. 
Taken all together, these results suggest that sophisticated skip connections work better and that networks can benefit from the increased context that multiscale dilations provide. 

\subsection{Transfer Learning for Speech Commands}

To evaluate whether convolutional networks applied to audio spectrograms were learning general properties of sound identification, we conducted a second set of experiments to classify speech commands.
In this second set, we train convolutional networks to classify speech commands and vary whether the network is pre-trained on the UrbanSound8K dataset and whether multiscale dilation is used as the network input. 
The classification task is broken into three separate tasks: discriminating between the commands \textit{left} vs. \textit{right}, discriminating between the 20 core speech commands, and discriminating between all 30 short utterance categories.
This was done to assess the quality of the learning.
In total, there were twelve conditions: freshly initialized vs pre-trained network, multiscale input vs no multiscale input, and three versions of the dataset.

\subsubsection{Training Details}

For all conditions, we used the DenseNet-121 architecture, the Stochastic Gradient Descent optimization algorithm with a momentum of 0.9, and the negative log likelihood loss criterion. 
To adapt the network in the pre-training conditions, the final linear layer was removed and replaced with one which had the correct number of output classes. 
For both the fresh initialization and pre-training conditions, the final linear layer had a learning rate of $0.005$ and weight decay of $1 \times 10^{-4}$ while the rest of the network had a learning rate of $0.001$ and no weight decay. 
Unlike the first set of experiments, we did not use an early stopping mechanism, but instead trained the models for exactly 100 epochs.

\begin{table}[!tp]
\centering
\begin{tabular}{ c|cc } 
& \multicolumn{2}{c}{No Multiscale} \\
   & Fresh Initialization & Pretrained \\
  \hline             
  \textit{left} vs. \textit{right} Subset    & 89.19 & 91.40 \\
  20 Command Terms     & 81.32 & 82.48 \\
  All 30 Terms         & 80.13 & 81.55 \\
\end{tabular}
\caption{For the Speech Commands dataset and the \textit{left} vs. \textit{right} subset, the classification performance for a model pre-trained on UrbanSound8k and a freshly initialized model are compared. These models did not have multiscale input.}
\label{table:gscresults_nomultiscale}
\end{table}

\begin{table}[!tp]
\centering
\begin{tabular}{ c|cc } 
 &  \multicolumn{2}{c}{Multiscale} \\
    & Fresh Initialization & Pretrained \\
  \hline              
  \textit{left} vs. \textit{right} Subset    & 88.54 & 95.32 \\
  20 Command Terms     & 82.22 & 85.52  \\
  All 30 Terms         & 82.11 & 84.35  \\
\end{tabular}
\caption{For the Speech Commands dataset and the \textit{left} vs. \textit{right} subset, the classification performance for a model pre-trained on UrbanSound8k and a freshly initialized model are compared.  These models additionally had multiscale input.}
\label{table:gscresults_multiscale}
\end{table}

\subsubsection{Results}

The results, shown in Tables~\ref{table:gscresults_nomultiscale} and \ref{table:gscresults_multiscale}, can be summarized with the following observations\footnote{Due to the recency of this dataset, these results are the only known results to the authors at the time of writing.}.
First, the pre-trained networks had increased performance over networks that started with freshly initialized parameters.
Next, the performance boost of all pre-trained convolutional networks is more pronounced with multiscale dilated convolutions as inputs.
Finally, it's an interesting point to note that despite being an order of magnitude larger than the UrbanSound8K dataset, the Google Speech Commands dataset still benefited from learning transfer using the pre-trained representations from an unrelated classification task.
This result is compelling because it suggests a strong interaction effect between the pre-training and multiscale dilated convolutions and warrants further investigation.

\subsection{Transfer Learning and Target Data Size}

A core lesson from the previous experiment is that a network pre-trained on environmental sound classification can be fine-tuned to perform well on classifying speech commands. While interesting, our goal was to study the contribution of pre-trained representations in transfer learning. In order to do so, we conducted an additional experiment that incrementally varied the amount of target data available for training. 
Evaluating models trained on limited target data gives a sense of how well the pre-trained network generalizes.

\subsubsection{Training Details}

\begin{figure}
\includegraphics[width=0.5\textwidth]{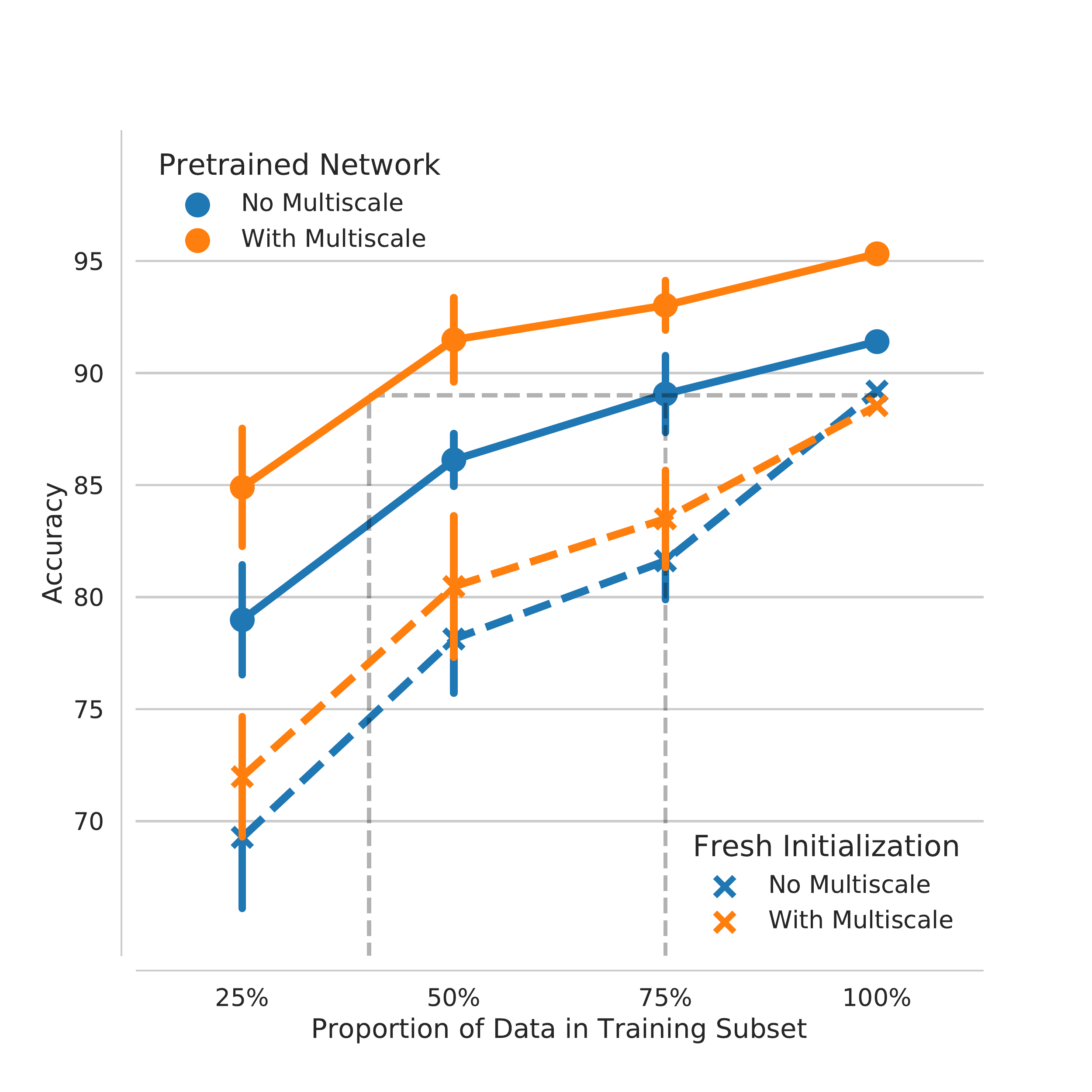}
\caption{For transfer learning, the DenseNet-121 architecture is trained on only the \textit{left} vs. \textit{right}  subset of the Google Speech Commands dataset using 25\%, 50\%, 75\% or 100\% of the data, with or without multiscale input, and with or without being pre-trained on UrbanSound8k.}
\label{fig:left-right-data-ablation}
\end{figure}

\begin{figure}
\includegraphics[width=0.49\textwidth]{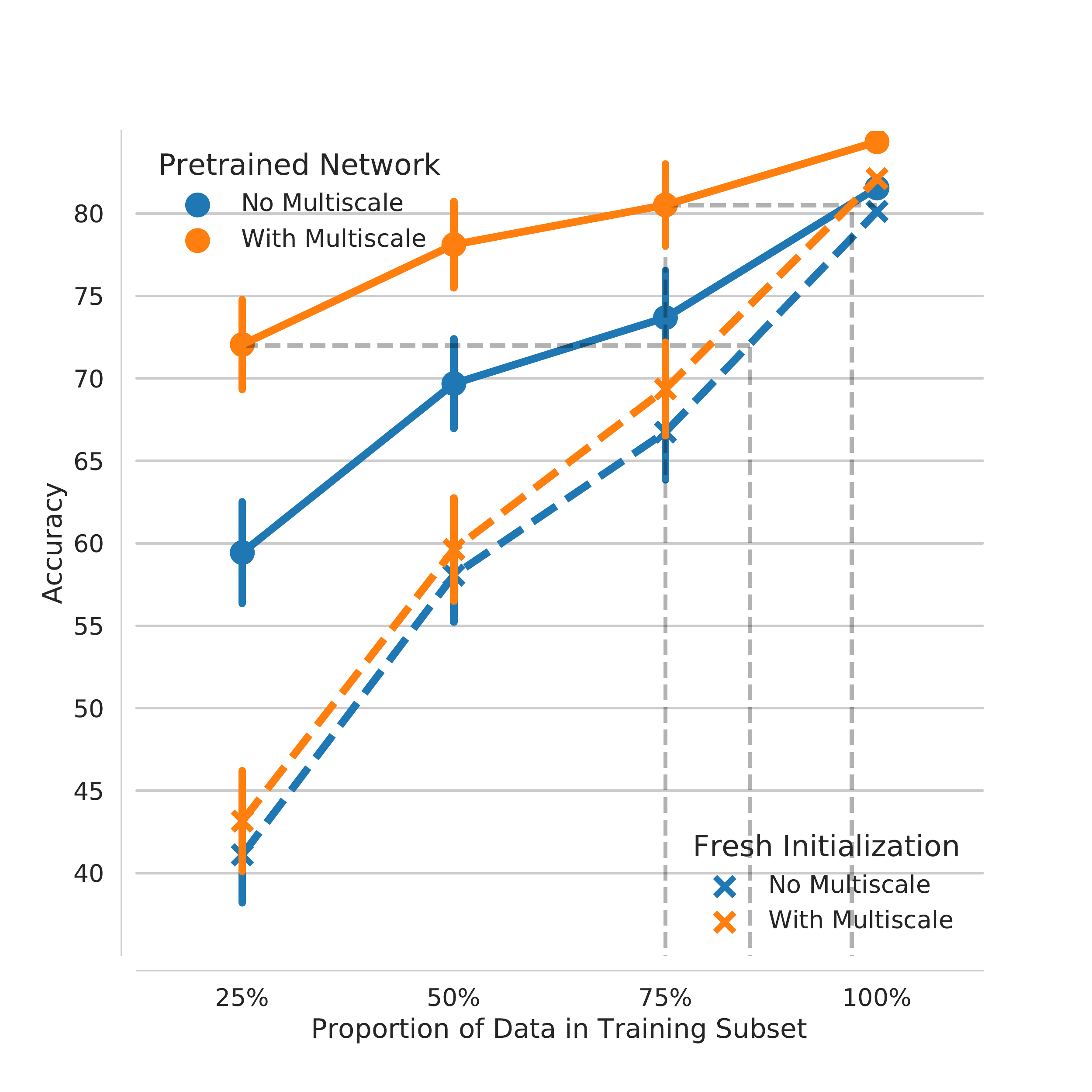}
\caption{For transfer learning, the DenseNet-121 architecture is trained on the whole Google Speech Commands dataset---\emph{20 commands and 10 non-commands}---using 25\%, 50\%, 75\%, or 100\% of the data, with or without multiscale input, and with or without being pre-trained on UrbanSound8k.}
\label{fig:all-data-ablation}
\end{figure}

In this experiment, we vary the amount of training data available to the network---using either 25\%, 50\%, 75\%, or 100\% of the training data---and test on the complete test set. 
For each subset size, we run five iterations and randomize the subset\footnote{Subset selection was done on the collection of data points for each class and each split of the dataset to ensure the ratio of data between classes remained constant.} on each iteration.
Given the training data subset, the remaining training details are kept the same as in the previous experiment.
We report the results of this setup on the \textit{left} vs. \textit{right}  subset of speech commands and on the whole dataset of 20 commands and 10 non-commands.

\subsubsection{Results}

The results are shown in Figure \ref{fig:left-right-data-ablation} for the \textit{left} vs. \textit{right}  subset and in Figure \ref{fig:all-data-ablation} for the whole dataset.  
There are several key take-away points.
First, for the \textit{left} vs. \textit{right} subset, using only 75\% of the training data, the pre-trained network obtained the same performance as the freshly initialized network with 100\% of the training data. 
The amount of training data drops further to 40\% when the multiscale input representation is used in conjunction with pre-training.
Second, for the whole dataset, using only 25\% of the data, the pre-trained network with multiscale inputs is able to achieve the accuracy as the freshly initialized network does with ~80\% of the data.  
The final take-away point is that the benefits of multiscale inputs are much lower in the freshly initialized networks.
This is strong evidence of the interaction between pre-training and multiscale inputs: using pre-trained multiscale input through dilated convolutions prominently increases the transfer capabilities of the network.

\section{Discussion}


In this paper, we have analyzed very deep convolutional networks classify audio spectrograms.
We have systematically enumerated several different convolutional network architectures and whether or not these networks were provided a multiscale input through dilated convolutions.
In this section, we provide a discussion over these enumerations to highlight lessons and outline exciting future directions.

\subsubsection{Lessons}

\begin{figure}
\includegraphics[width=0.5\textwidth]{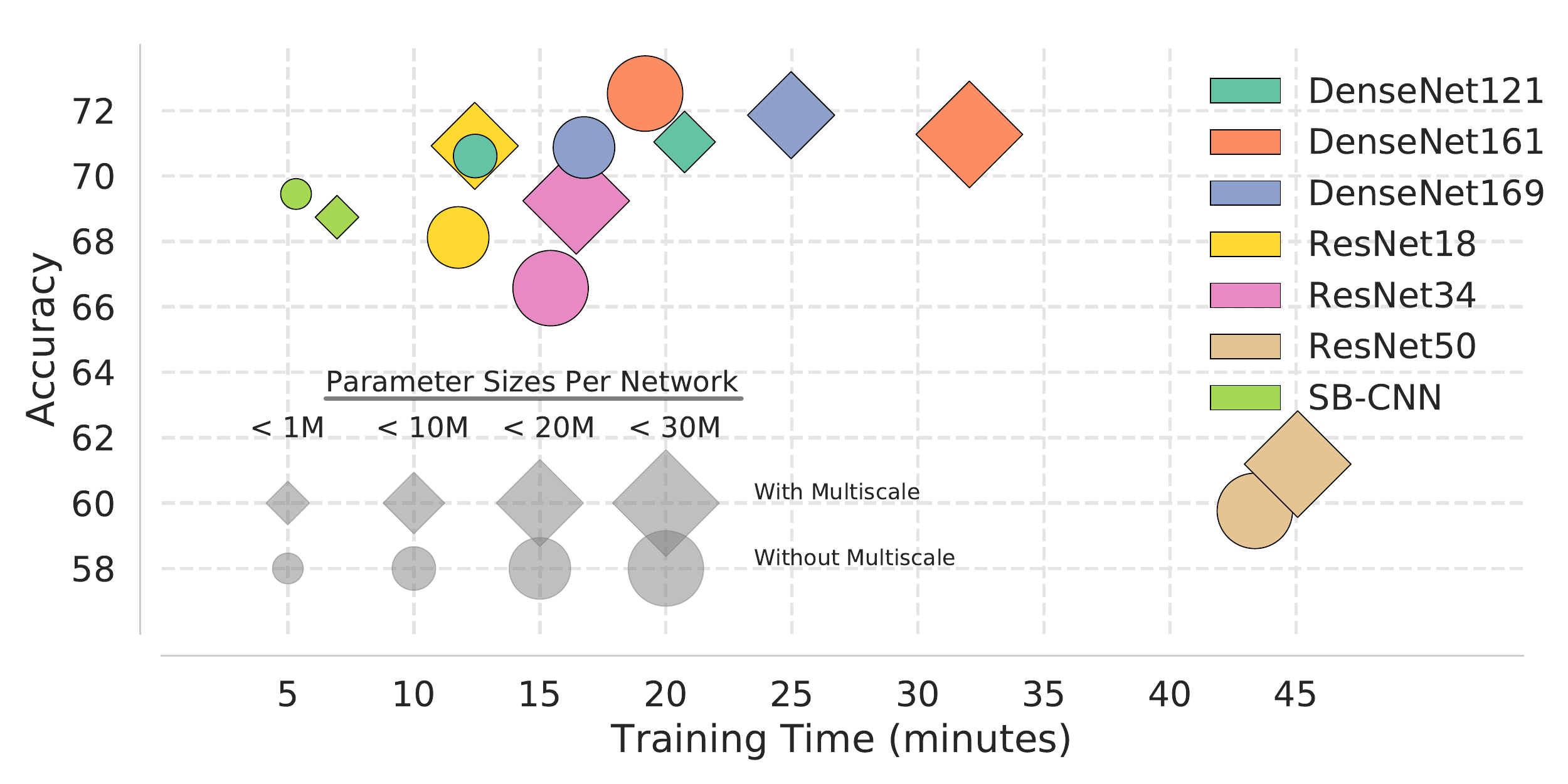}
\caption{The classification accuracies for the convolutional network architectures---both with and without multiscale inputs using dilated kernel---are shown as a function of model size (number of parameters) and average training time.}
\label{fig:size_perf_time}
\end{figure}

The first lesson that we can draw concerns which convolutional network architecture to use when learning from labeled audio spectrograms. 
In Figure \ref{fig:size_perf_time}, we plot the various network architectures, with and without multiscale, and annotated with the number of parameters as a function of their training time and accuracy. 
Overall, the evidence suggests that DenseNet architectures are well suited to this domain. 
Further, the results highlight a promising pattern of multiscale positively impacting performance.

The next lesson that we can draw focuses on the role of transfer learning between environmental sounds and speech commands. 
Not only were the pre-trained networks able to obtain higher accuracies with smaller subsets of the target data, but this pattern was magnified for pre-trained networks which had the multiscale input using dilated convolutions. 
This suggests that the multiscale dilated convolutions could be learning something important about sound identification that transfers well to speech command classification. 
Additionally, the results offers a useful insight into practical considerations that one can use less target data if the network has been pre-trained.
This study does not, however, fully characterize what properties of the source data are necessary for successful transfer representations well for audio classifications tasks in convolutional networks.
Perhaps it is the nature of environmental sounds that allowed the network to capture generalizable patterns, and that less diverse audio will not work as well as the source dataset in the transfer setting. 
This needs further investigation.

\subsubsection{Future Work}

The study presented in this paper serves as a starting point from which several intriguing approaches could be pursued.
Our model performances, while near state of the art, are not at the level of human performance\footnote{Reported as 82\% on 4 second environmental sound clips~\cite{Chu09}.}.
One promising approach is to evaluate how data augmentation interacts with our findings on pre-training and multiscale input with dilated kernels. 

Another encouraging follow-up is to further investigate the impact of multiscale inputs.
While this study observed that the multiscale input using dilated convolutions improved classification performance on freshly initialized networks and compounded the effectiveness of transfer learning, there are potential conflations that should be carefully studied and ruled out.
In general, though, the promising results of multiscale inputs using dilated convolutions suggest that they can be combined with other techniques and warrants further study. 

\section{Conclusion}

In this work, we have presented a study of convolutional network architectures applied to classifying the audio spectrograms of the UrbanSound8k and Google Speech Commands dataset.
Our contributions are an exposition into the relationship between convolutional network architectures and audio spectrograms, a novel multiscale input using dilated convolutions, and an examination of how well learning can transfer from an environment sounds dataset to a speech commands dataset.

We conclude by summarizing these contributions in five findings. 
First, we find that DenseNets perform very well on audio spectrogram classification when compared to a baseline network and ResNet.
This confirms a trend in Computer Vision correlates performance in various vision tasks with increasingly deep and sophisticated convolutional network architectures. 
Second, our novel use of dilated convolutions for multiscale inputs resulted in increased performance on the classification tasks.

The next three findings are centered on the transfer learning experiments.
To begin, our third finding is that convolutional networks pre-trained on environmental sound classification out-performed freshly initialized convolutional networks on the task of classifying speech commands. 
As a consequence, our fourth finding is that this gap in performance means we can obtain the same classification accuracy with less training data. 
Finally, our fifth finding is that the previous two findings are even stronger when multiscale inputs through dilated convolutions are employed.
In other words, pre-training on environmental sounds with a convolutional network that utilizes multiscale inputs through dilated convolutions can dramatically increase classification accuracy with a fraction of the data. 

Through this study, we have evaluated a series of convolutional network architectures and different modeling choices on audio spectrograms.
Further, we have demonstrated a relationship between the kinds of representations needed for recognizing environmental sounds and for recognizing speech commands. 
Moving forward, there are many promising directions which can further unify audio event identification for both human speech and ambient environmental sounds.

\bibliography{urbansound}
\bibliographystyle{aaai}

\end{document}